\documentstyle[12pt,epsfig,amsmath,amssymb,pstricks]{article}
\begin{document}

\title{\bf Hilbert-space factorization is a limited and expensive information-processing resource}

\author{{Chris Fields}\\ \\
{\it 528 Zinnia Court}\\
{\it Sonoma, CA 95476 USA}\\ \\
{fieldsres@gmail.com}}
\maketitle

\begin{abstract}
By taking the need for quantum reference frames into account, it is shown that Hilbert-space factorization is a dissipative process requiring on the order of $kT$ to reduce by one bit an observer's uncertainty in the provenance of a classically-recorded observational outcome.  This cost is neglected in standard treatments of decoherence that assume that observational outcomes are obtained by interacting with a collection of degrees of freedom identified \textit{a priori}.  Treating this cost explicitly leads to a natural measure of the probability of any particular quantum reference frame.
\end{abstract}


\section{Introduction}

Wojciech Zurek opened his 1998 ``rough guide'' to decoherence by noting that while physics requires a universe split into systems, ``it is far from clear how one can define systems given an overall Hilbert space `of everything' and the total Hamiltonian'' (\cite{zurek:98} p. 1794).  In ``Consciousness as a State of Matter'' \cite{tegmark:14}, Max Tegmark takes up this Hilbert-space factorization problem from an unusual direction: an analysis of the essential characteristics of an observer loosely based on Tononi's \cite{tononi:08} information-integration theory of human consciousness.  The most important of these characteristics proves to be the ``Principle of Utility'' - the idea that consciousness is good for something, namely prediction.  This idea, which echoes the motivation for Zurek's well-known ``predictability seive'' method for identifying einselected and hence environmentally-encoded pointer states, at least hints at an anthropic solution to the problem of defining preferred Hilbert-space factorizations: ``if we find that useful consciousness can only exist given certain strict requirements on the quantum factorization, then this could explain why we perceive a factorization satisfying these requirements'' (p. 29).

This Comment focuses on the first of the three problems Tegmark highlights as standing in the way of a full solution to the Hilbert-space factorization problem: what he calls the ``chicken-and-egg'' problem of ``what should we determine first: the state or the factorization?'' (\cite{tegmark:14} p. 28).  An examination of this problem leads to an implicit assumption: the assumption that the state of a quantum degree of freedom can be queried by interacting with \textit{only} that degree of freedom.  This assumption is rendered in information-processing terms as the assumption that Hilbert-space factorization is a freely-available resource.  It is then shown that this assumption is false: Hilbert-space factorization is not free.  It is limited and expensive.  Recognizing that Hilbert-space factorization is expensive sheds new light on the third problem Tegmark highlights, that of the emergence of time in unitary dynamics.  It suggests, in particular, that the \textit{useful} Hilbert-space factorizations are precisely those that allow time to be measured, as it is in these factorizations that observers are able locally to reduce entropy.

\section{Factorization is a resource}

With how many degrees of freedom must Schr\"{o}dinger interact to determine whether his cat is dead or alive?  The standard answer is one: the cat is a qubit $\psi_{cat}$ in some state $|\psi_{cat}\rangle = \alpha |dead\rangle + \beta |alive\rangle$ and Schr\"{o}dinger needs merely to measure this qubit in the $\{ |dead\rangle, |alive\rangle \}$ basis.  This standard answer is encapulated in the standard picture of decoherence, here adapted from \cite{tegmark:12}: the total Hilbert space $\mathcal{H}_{\mathbf{U}}$ is factored into a 3-component tensor product $\mathcal{H}_{\mathbf{U}} = \mathcal{H}_{\mathbf{S}} \otimes \mathcal{H}_{\mathbf{O}} \otimes \mathcal{H}_{\mathbf{E}}$ where the ``system'' component $\mathcal{H}_{\mathbf{S}}$ contains \textit{only} the degrees of freedom of interest to the observer (the qubit $\psi_{cat}$ in Schr\"{o}dinger's case), the ``observer'' component $\mathcal{H}_{\mathbf{O}}$ contains \textit{only} the degrees of freedom employed in the observer's representation of the observational outcome (e.g. Schr\"{o}dinger's representation of the pointer values ``\textit{dead}'' and ``\textit{alive}'') and the ``environment'' component $\mathcal{H}_{\mathbf{E}}$ contains all other degrees of freedom in the universe, and in particular, all other degrees of freedom of both Schr\"odinger and his cat.  With this factorization, the overall Hamiltonian is:

\begin{equation}
H_{\mathbf{U}} = \mathit{H}_{\mathbf{S}} + \mathit{H}_{\mathbf{O}} + \mathit{H}_{\mathbf{E}} + \mathit{H}_{\mathbf{SO}} + \mathit{H}_{\mathbf{SE}} + \mathit{H}_{\mathbf{OE}} + \mathit{H}_{\mathbf{SOE}},
\end{equation}

where the first three terms represent the internal dynamics of $\mathbf{S}$, $\mathbf{O}$ and $\mathbf{E}$
respectively, $\mathit{H}_{\mathbf{SO}}$ represents the measurement interaction, $\mathit{H}_{\mathbf{SE}}$ and $\mathit{H}_{\mathbf{OE}}$ represent environmental decoherence acting on $\mathbf{S}$ and $\mathbf{O}$
respectively, and $\mathit{H}_{\mathbf{SOE}}$ represents any residual interaction and is never discussed further.  Figure 1 illustrates this situation.

\begin{center}

\psset{xunit=1cm,yunit=1cm}
\begin{pspicture}(0,0)(8,10)

\pswedge(4.4,6.5){2.5}{-30}{90}
\put(4.8,7.5){$System$}
\put(5.2,7){$\mathcal{H}_{\mathbf{S}}$}
\pswedge(3.6,6.5){2.5}{90}{210}
\put(1.6,7.5){$Observer$}
\put(2.1,7){$\mathcal{H}_{\mathbf{O}}$}
\pswedge(4,5.9){2.5}{210}{330}
\put(2.8,4.4){$Environment$}
\put(3.7,4.9){$\mathcal{H}_{\mathbf{E}}$}
\psline{<->}(3.7,7.8)(4.3,7.8)
\put(3.6,8){$\mathit{H}_{\mathbf{SO}}$}
\psline{<->}(2.7,5.9)(3,5.4)
\put(1.8,5.2){$\mathit{H}_{\mathbf{OE}}$}
\psline{<->}(5.3,5.9)(5,5.4)
\put(5.4,5.3){$\mathit{H}_{\mathbf{SE}}$}

\put(0.5,1.5){\textit{Fig. 1:}  Decoherence, adapted from \cite{tegmark:12},}
\put(0.5,1){Fig. 2.  The residual 3-way interaction}
\put(0.5,0.5){$\mathit{H}_{\mathbf{SOE}}$ is as usual taken to be negligible.}
\end{pspicture}

\end{center}

As emphasized in \cite{tegmark:12}, the degrees of freedom of the environment component $\mathbf{E}$ are always traced over: this trace operation is what enables decoherence by rendering any $\mathbf{S}$ - $\mathbf{E}$ or $\mathbf{O}$ - $\mathbf{E}$ entanglement invisible to $\mathbf{O}$.  The observer can, therefore, have no knowledge of $H_{\mathbf{E}}$; in practice, $\mathbf{E}$ is treated as a classical heat bath and the representations of $\mathit{H}_{\mathbf{SE}}$ and $\mathit{H}_{\mathbf{OE}}$ are chosen to reflect this (e.g. \cite{joos-zeh:85, zurek:03, schloss:07}).  Decoherence of $\mathbf{S}$ and $\mathbf{O}$ by $\mathit{H}_{\mathbf{SE}}$ and $\mathit{H}_{\mathbf{OE}}$ respectively is taken to be effectively instantaneous, so that relative states of $\mathbf{S}$ and $\mathbf{E}$ can be considered to be classically correlated, i.e. all observational outcomes can be considered to have definite values.

With this representation of decoherence, the chicken-and-egg problem can be stated clearly: if the factorization is given, pointer states of $\mathbf{S}$ follow by einselection, but \textit{finding} an optimal factorization requires knowing something about the current state of $\mathbf{S}$.  Tegmark summarizes this second aspect of the problem as follows:

\begin{quote}
However, to find the best factorization, we need information about the state. A clock is a highly autonomous system if we factor the Hilbert space so that the first factor corresponds to the spatial volume containing the clock, but if the state were different such that the clock were somewhere else, we should factor out a different volume.

\begin{flushright}
\cite{tegmark:14} p. 28
\end{flushright}
\end{quote}

This statement is interesting, in part, because the position of the spatial volume containing a clock is seldom considered to be the clock's pointer state: it is a degree of freedom that would ordinarily be relegated to the ``environment'' component in Fig. 1.  But let us suppose that the spatial volume of interest is the particular spatial volume containing just the hand of the clock, i.e. the volume for which the positional degree of freedom generally \textit{is} regarded as the clock's pointer state.  Does interacting only with this pointer degree of freedom allow an observer to determine the time?  Clearly not: the observer must also determine both the position of the clock's face and its angular orientation relative to some fixed point outside of the clock.  Unless the clock's face is also observed, the position of the clock's hand yields no information about the time at all; indeed nothing about a ``bare'' hand indicates even that it is the hand of a clock.  The clock's face serves here is a quantum reference frame \cite{bartlett:07} without which the time cannot be determined from observations of the hand, and which must be shared by any two observers both using the hand position to determine the time.  Hence Tegmark's chicken-and-egg problem is more interesting than it at first appears: in order to use information about the state of $\mathbf{S}$ to find an appropriate factorization, an observer needs to employ a \textit{different} factorization, one that picks out a quantum reference frame that enables measurements of the state of $\mathbf{S}$ to convey useful information.  To determine his cat's health, in other words, Schr\"{o}dinger cannot simply measure the qubit $\psi_{cat}$.  First he has to \textit{pick out his cat} from among the other degrees of freedom in the universe, and that requires observing degrees of freedom - coat color, for example, or number of legs - other than its health.

This situation generalizes, and it does so in a way that is instantly clear from the perspective of experimental design.  Measuring any quantum degree of freedom, or any collection of quantum degrees of freedom, that compose a system of interest $\mathbf{S}$ requires the prior identification of some other set of quantum degrees of freedom $\mathbf{R}$ that compose a quantum reference frame, one typically called an ``apparatus.''  No measurement can get off the ground until the observer has located the apparatus with which the observation is to be made, and then ``prepared'' the apparatus to assure that it is working properly.  This preparation step requires the observer to have some knowledge of the dynamics $H_{\mathbf{R}}$ that govern the behavior of $\mathbf{R}$; hence $\mathbf{R}$ \textit{cannot} be traced over.  From the perspective of Fig. 1, this apparatus $\mathbf{R}$ is part of $\mathbf{E}$; indeed its ability to function as a quantum reference frame requires that it be part of neither $\mathbf{S}$ nor $\mathbf{O}$.  Thus while any particular environmental degrees of freedom can be incorporated by stipulation into $\mathbf{S}$ by adopting a revised factorization, some degrees of freedom must remain in $\mathbf{E}$ to serve as a quantum reference frame.  One could, for example, bundle all the degrees of freedom of the LHC into $\mathbf{S}$, but without the degrees of freedom of the surrounding CERN infrastructure, the internet, or the local geography of Geneva to serve as a reference frame, these degrees of freedom would not be identifiable as those of the LHC, and hence would provide no useful information.  An observer who pursues von Neumann's chain to the bitter end and bundles \textit{all} degrees of freedom except her own subjective representation of the outcome into $\mathbf{S}$ - and hence eliminates $\mathbf{E}$ altogether from Fig. 1 - becomes a locus of ``pure experience'' \cite{fuchs:10} indistinguishable from a Boltzmann brain. 

The observer's carving-out of some subset of environmental degrees of freedom to serve as a quantum reference frame $\mathbf{R}$ is illustrated in Fig. 2.  Here $\mathbf{E^{\prime}}$ is the traced-over remainder of the environment, so $\mathcal{H}_{\mathbf{R}} \otimes \mathcal{H}_{\mathbf{E^{\prime}}} = \mathcal{H}_{\mathbf{E}}$.  All terms in Eqn. 1 referring to $\mathbf{E}$ may be expanded:

\begin{align}
&H_{\mathbf{E}} = \mathit{H}_{\mathbf{R}} + \mathit{H}_{\mathbf{E^{\prime}}} + H_{\mathbf{RE^{\prime}}}; \\ &\mathit{H}_{\mathbf{OE}} = \mathit{H}_{\mathbf{OR}} + \mathit{H}_{\mathbf{OE^{\prime}}}; \\ &\mathit{H}_{\mathbf{SE}} = \mathit{H}_{\mathbf{SR}} + \mathit{H}_{\mathbf{SE^{\prime}}};
\end{align}

where $H_{\mathbf{E^{\prime}}}$ represents the internal dynamics of the environmental remainder $\mathbf{E^{\prime}}$ and $H_{\mathbf{RE^{\prime}}}$ represents its interaction with $\mathbf{R}$.  Neglecting higher-order terms as before, the qualitative consequences of this carving out of $\mathbf{R}$ from $\mathbf{E}$ can be examined in the relevant two limits.  As $\mathbf{E}^{\prime} \rightarrow \mathbf{E}$ the observer interacts with a ``bare pointer'' unrelated to any reference frame as in Fig. 1.  As noted earlier in the case of a clock, the state of such a bare pointer confers no useful information to $\mathbf{O}$.  As $\mathbf{R} \rightarrow \mathbf{E}$, the reference frame expands to include the entire rest of the universe.  Bearing in mind that $\mathbf{R}$ only serves as a reference frame for an observer capable of monitoring its degrees of freedom - monitoring the position and orientation of the clock's face while telling the time, for example - it is clear that this limit is achievable only if $\mathbf{O}$ has sufficient encoding capacity to monitor the entire state of $\mathbf{E}$, i.e. only if $\mathbf{O} \gg \mathbf{E}$.  This condition, however, is just a re-labelling of Fig. 1 that swaps $\mathbf{O}$ and $\mathbf{E}$.  It is the perspective of the ``environment as witness'' \cite{zurek:04, zurek:05} that not only imposes decoherence but continuously monitors and hence encodes the states of all other degrees of freedom.  The situation of interest, therefore, lies between these limits: it is the case of a ``reasonable'' quantum reference frame that enables measurements of $\mathbf{S}$ to convey useful information without requiring that $\mathbf{O}$ have sufficient encoding capacity to monitor the entire rest of the universe.

\begin{center}

\psset{xunit=1cm,yunit=1cm}
\begin{pspicture}(0,0)(8,10)

\pswedge(4.4,6.5){2.5}{-30}{90}
\put(4.8,7.5){$System$}
\put(5.2,7){$\mathcal{H}_{\mathbf{S}}$}
\pswedge(3.6,6.5){2.5}{90}{210}
\put(1.6,7.5){$Observer$}
\put(2.1,7){$\mathcal{H}_{\mathbf{O}}$}
\pswedge(4,5.9){2.5}{210}{330}
\put(2.8,4){$Environment^{\prime}$}
\put(2.2,4.5){$\mathcal{H}_{\mathbf{E}^{\prime}}$}
\put(3.5,4.7){$QRF$}
\put(3.6,5.3){$\mathcal{H}_{\mathbf{R}}$}
\psarc[linestyle=dashed](4,5.9){1.5}{210}{330}

\psline{<->}(3.7,7.8)(4.3,7.8)
\put(3.6,8){$\mathit{H}_{\mathbf{SO}}$}
\psline{<->}(2.7,5.9)(3,5.4)
\put(1,4.8){$\mathit{H}_{\mathbf{OE^{\prime}}}$}
\psline{<->}(5.3,5.9)(5,5.4)
\put(6,4.8){$\mathit{H}_{\mathbf{SE^{\prime}}}$}

\psline{<->}(2,5.5)(2.3,5)
\put(2.9,5.8){$\mathit{H}_{\mathbf{OR}}$}
\put(4.2,5.8){$\mathit{H}_{\mathbf{SR}}$}
\psline{<->}(5.7,5)(6,5.5)

\put(0.5,2){\textit{Fig. 1:}  Decoherence, with a quantum}
\put(0.5,1.5){reference frame ($QRF$) as part of the}
\put(0.5,1){environment.  The interaction $H_{\mathbf{RE^{\prime}}}$ is}
\put(0.5,0.5){defined at the dashed $\mathbf{R}$-$\mathbf{E}^{\prime}$ boundary.} 
\end{pspicture}

\end{center}

As emphasized in \cite{bartlett:07}, quantum reference frames are \textit{resources} that observers must provide if they are to extract useful information from the world.  Obtaining a quantum reference frame requires having a factorization with which to distinguish it from the rest of the environment; hence such factorizations must also be considered resources.  The question underlying Tegmark's chicken-and-egg problem is, therefore, the question of what these resources cost.  If factorizations are \textit{free} resources, they can be treated as \textit{given} and einselection does the rest; hence if factorizations are free, the chicken-and-egg problem presents no difficulty in practice.  

\section{Factorization is not free}

If one considers the factorization that picks out the LHC as a resource, one receives a strong hint that such resources are not free.  Even constructing a meter stick requires energy.  To see that this cost is general, consider the process of observation from the perspective of an observer equipped only with a single observable - say $\hat{x}$ - and an effectively-classical memory.  The observer operates on the world with $\hat{x}$, and records a succession of position eigenvalues in memory.  As in the case of a human infant exploring her environment, or of an ``infant robot'' such as the iCub \cite{sandini:07}, rendering these recorded eigenvalues useful for making predictions requires solving an inverse problem.  In particular, it requires determining when two eigenvalues of $\hat{x}$ were obtained from the \textit{same} system $\mathbf{S}$.  In a more careful formalism in which observables such as $\hat{x}$ are defined as automorphisms on Hilbert spaces corresponding to particular collections of degrees of freedom, i.e. as $\hat{x}_{\mathbf{S1}}$, $\hat{x}_{\mathbf{S2}}$, etc. acting on Hilbert spaces $\mathcal{H}_{\mathbf{S1}}$, $\mathcal{H}_{\mathbf{S2}}$, etc., this inverse problem is the problem of determining when two recorded eigenvalues were produced by the same factorization-specific observable.  Call this inverse problem the ``quantum system identification problem'' or QSIP.  The QSIP is the problem of constructing a factorization given one or more classically-recorded outcome values; hence it corresponds to the ``egg'' side of Tegmark's chicken-and-egg problem, the side on which factorizations are not assumed to be freely given.  The fact that the outcome values have, by assumption, been classically recorded requires that the constructed factorization supports decoherence, i.e. that $\mathbf{E}$ (Fig. 1) or $\mathbf{E}^{\prime}$ (Fig. 2) are both sufficiently large and unobserved by $\mathbf{O}$ and hence traced over.

In the language of quantum Darwinism \cite{zurek:09}, the QSIP is the problem of determining whether two environmental encodings of a pointer state are redundant encodings of the \textit{same} pointer state produced by environmental interactions with a single system, or encodings of \textit{different} pointer states produced by environmental interactions with two different systems.  Note that in quantum Darwinism, each observer $\mathbf{O}_{\mathit{i}}$ has an effective quantum reference frame: the fragment $\mathbf{F}_{\mathit{i}}$ of the environment in which that observer is confined.  These fragments are by definition non-overlapping, so observers cannot \textit{share} a reference frame; hence they can communicate only using fungible \cite{bartlett:07} information.  As an observer has no observational access to the boundary separating their fragment $\mathbf{F}_{\mathit{i}}$ from the rest of $\mathbf{E}$ - the dashed boundary in Fig. 2 - they have no way of assuring that $\mathbf{F}_{\mathit{i}}$ remains the \textit{same} reference frame from one observation to the next.  Hence any observer who makes multiple observations at different times faces the QSIP, as does any pair of observers who wish to compare observations made at the same time.  Anyone who has mistaken a merely similar car for their own in a parking lot, or has wondered whether they and a conversation partner are talking about the same thing has encountered the QSIP.

That the QSIP is unsolvable in principle by finite sequences of non-destructive measurements was proven in the early days of cybernetics: not even classical systems can be unambiguously identified by such means \cite{ashby:56, moore:56}.  Destructive measurements are similarly insufficient \cite{fields:10, fields:11}; to see this, it suffices to note that an observer seeking to identify the source of an environmental encoding after a destructive manipulation of some system is in exactly the same situation as they were in when attempting to identify the source of the encoding before the manipulation.  The reason for this difficulty is obvious: determining the provenance of an environmental encoding requires access to the Hamiltonian of the environment, and observational access to this Hamiltonian is removed by the trace over $\mathbf{E}$ that is required for decoherence.  An observer confined to an environmental fragment $\mathbf{R} = \mathbf{F}_{\mathit{i}}$ must, in particular, have access to $H_{\mathbf{E}} = \mathit{H}_{\mathbf{R}} + \mathit{H}_{\mathbf{E^{\prime}}} + H_{\mathbf{RE^{\prime}}}$, which both confinement to $\mathbf{F}_{\mathit{i}}$ and the required trace over $\mathbf{E^{\prime}}$ prevent.  Thus while decoherence makes pointer states useful for prediction by rendering them recordable in environmental fragments that serve as classical memories, it undercuts this utility by rendering their provenance uncertain.

An observer who has traced over the environment knows only that she is currently interacting with some system or other, hence her uncertainty about the source of a recorded outcome value can be represented as an entropy (in bits):

\begin{equation}
S_{F} = k log_{2} \Omega_{F},
\end{equation}

where $\Omega_{F}$ counts the number of distinct factorizations of $\mathcal{H}_{\mathbf{U}}$ that could have produced the recorded value.  For an observer who has recorded a binary string but has traced over $\mathbf{E}$, for example, $\Omega_{F}$ is the number of qubits in the universe.  For an observer equipped only with a meter stick, $\Omega_{F}$ counts the number of macroscopic objects.  From the observer's perspective, $S_{F}$ is an entropy associated with $\mathbf{E}$; it is the Hamiltonian $H_{\mathbf{E}}$ that determines which degree(s) of freedom the observer is interacting with.  Again this is clearest from the perspective of quantum Darwinism: if observers obtain information by examining encodings of distant pointer states in their local environmental fragments, it is $H_{\mathbf{E}}$ that determines what pointer states have been encoded.  Thus while the standard environmental entropy $S_{\mathbf{E}}$ measures an observer's uncertainty about the \textit{state} of the environment \cite{tegmark:12}, the factorization entropy $S_{\mathbf{F}}$ effectively measures the observer's uncertainty about the environment's \textit{dynamics}.  This distinction allows a clean re-statement of the central result of \cite{tegmark:12}: while decoherence considerations show that only a relatively small number of observations are required to determine the state of one's local Hubble volume - in \cite{tegmark:12}, to determine whether it is post-inflationary and hence able to support life - under conditions of decoherence these observations can reveal nothing about the \textit{dynamics} of one's local Hubble volume.  Decoherence allows $S_{\mathbf{E}}$ to be low, but forces $S_{\mathbf{F}}$ to be high.  This is significant, because $S_{\mathbf{F}}$ measures uncertainty about \textit{which} collections of degrees of freedom $H_{\mathbf{U}}$ has carried into the local Hubble volume.  It is this uncertainty that drives the measure problem.

The factorization entropy $S_{F}$ is the hidden cost that observers pay for decoherence, the cost of tracing over $\mathbf{E}$.  The common assumption that observers know the dynamical provenance of their observations and hence know which degrees of freedom they are observing ignores this hidden cost.  Real observers, however, cannot ignore $S_{F}$.  They reduce $S_{F}$ by expending energy to locally control the Hamiltonian of the environment.  Observers exert this control by organizing environmental degrees of freedom into quantum reference frames that are continuously monitored and hence controlled by the quantum Zeno effect.  The contributions of environmental degrees of freedom that are monitored in this way can be subtracted from $\Omega_{F}$, reducing uncertainty about the provenance of recorded outcome values.  As quantum reference frames tend to be macroscopic objects - meter sticks, voltmeters, computers, laboratory buildings, orbiting telescopes, particle accelerators - each bit of uncertainty reduction costs on the order of $kT$.  Observation is, therefore, a bulk dissipative process: observers are not just reducing the entropy of a few degrees of freedom contained in $\mathbf{S}$ as described in standard models of decoherence (e.g. \cite{tegmark:12}) but are reducing the entropy of $\mathbf{E}$ as well.  It is this requirement for macroscopic entropy reduction in the environment that explains the macroscopic Joules-per-bit cost of obtaining useful classical information even from single qubits.

In addition to its energetic cost, monitoring a quantum reference frame requires a memory expenditure of at least two bits for every environmental degree of freedom monitored - one bit for the current outcome value plus one for the previous outcome value - plus sufficient processing cycles to compare the current to the previous outcome value.  An observer's ability to reduce $S_{F}$ is, therefore, strictly limited by the observer's access to both memory and processing cycles.  That sufficiently creative observers offload some of these memory and computing requirements onto inexpensively-monitored components of the environment is, therefore, not surprising.

Tegmark concludes his discussion of the chicken-and-egg problem by suggesting that ``we need a criterion for identifying conscious observers, and then a prescription that determines which factorization each of them will perceive'' (\cite{tegmark:14} p. 28-29).  Consideration of the cost of factorization suggests a heuristic: look around for chunks of the environment that aren't moving.  Such low-entropy regions are probably being used by observers as quantum reference frames, and being held at low entropy by the quantum Zeno effect.  Examining the structure of these quantum reference frames will at least provide useful hints about what factorizations the local observers perceive.

\section{The emergence of time}

In order to be useful, a quantum reference frame must be re-identifiable as the \textit{same} quantum reference frame over the course of multiple observations; as noted above, it is this requirement that renders an otherwise-uncharacterized fragment of the local environment useless as a reference frame.  This requirement of re-identifiability applies, first and foremost, to the quantum reference frame essential for any observation: the observer's memory.  An observer who cannot re-identify her own memory, and hence cannot distinguish previous outcome values retrieved from memory from current outcome values received from the world is critically handicapped, as the plight of patients with damage to the ``reality monitoring system'' in Brodmann Area 10 of the prefrontal cortex demonstrates \cite{simons:08}.  A memory that is re-identifiable imposes a partial order on outcome values, even if it is only the partial order distinguishing ``previous'' from ``current'' outcomes.  Even the most primitive re-identifiable memory, in other words, imposes \textit{subjective} time.

If an encoding that imposes subjective time is necessary for useful factorizations, is it sufficient?  A factorization in which an observer continues to experience subjective time is at least a factorization in which that observer still exists.  Once observers are recognized as energy-dissipating systems, this becomes non-trivial: it requires the joint action of the Hamiltonians $H_{\mathbf{OR}}$, $H_{\mathbf{O}}$, and $H_{\mathbf{OE^{\prime}}}$ to transfer net energy from $\mathbf{E^{\prime}}$ to $\mathbf{R}$ and hence net entropy from $\mathbf{R}$ to $\mathbf{E^{\prime}}$, and to continue doing so over macroscopic (subjective) time.  An observer $\mathbf{O}$ can, therefore, be characterized by a net lifetime contribution of $\Delta S_{F} (\mathbf{O})$ to the construction and maintenance of quantum reference frames.  This is perhaps the best definition of ``utility'' that can be hoped for.

\section{Conclusion}

By examining Tegmark's ``chicken-and-egg'' problem of ``what should we determine first: the state or the factorization?'' \cite{tegmark:14}, we have seen that the answer is ``neither.''  What we should determine first, as Tegmark himself remarks in discussing this problem, is the identity of the observer.  Observers are dissipative systems that organize the environment around them into quantum reference frames, which they then monitor in an attempt to reduce their perceived factorization entropy $S_{F}$.  This activity constructs a factorization, or rather an equivalence class of factorizations that support the monitored quantum reference frames and thus have reduced entropy.  Within this equivalence class of factorizations, the provenance of any given observational outcome is still uncertain, but it is as certain as the observer can make it.

Viewing Hilbert-space factorizations as expensive, observer-supplied resources turns the usual notion of the ``emergence of classicality'' on its head.  No physical process, least of all decoherence, makes classical information freely available to observers.  A traced-over environment, in particular, is not a \textit{useful} witness, as the provenance of its encodings and hence their redundancy can never be determined \cite{fields:14}.  Creating classical information takes energy, on the order of $kT$ per bit, and this energy must be expended by observers.  What ``emerges'' from the action of $H_{\mathbf{U}}$ are dissipative systems that function as observers by transferring net energy in their local neighborhoods of $\mathcal{H}_{\mathbf{U}}$.  Classical information in the form of local net entropy transfer is a consequence of this activity.

A Hubble volume, from this perspective, is a large and incompletely-monitored quantum reference frame.  One may suggest, therefore, that the probability $P_{V}$ of a Hubble volume is inversely proportional to the total entropy transfer required to maintain it.  Assuming that the average observer contributes a lifetime utility $\Delta S_{F} (\mathbf{O}_{\mathit{av}})$, 

\begin{equation}
P_{V} \sim N_{\mathbf{O}}^{-1}
\end{equation}

where $N_{\mathbf{O}}$ is the number of observers inhabiting the volume.  With this measure, a Hubble volume in which every source of free energy is thickly colonized by observers is unusual.  One like ours, in which the majority of stars at least appear to lack habitable planets, is rather ordinary.


\begin{thebibliography}{99}

\bibitem{zurek:98}
W. H. Zurek.
\newblock Decoherence, einselection and the existential interpretation (the rough guide).
\newblock {\em Phil. Trans. Royal Soc. London}, 356: 1793-1821 (1998).
 
\bibitem{tegmark:14}
M. Tegmark.
\newblock Consciousness as a state of matter.
\newblock arXiv:1401.1219v1 [quant-ph], 2014.

\bibitem{tononi:08}
G. Tononi.
\newblock Consciousness as integrated information: A provisional manifesto.
\newblock {\em Biological Bulletin}, 215: 216-242, 2008.

\bibitem{tegmark:12}
M. Tegmark.
\newblock How unitary cosmology generalizes thermodynamics and solves the inflationary entropy problem.
\newblock {\em Phys. Rev. D}, 85: 123517, 2012.

\bibitem{joos-zeh:85}
E.~Joos and D.~Zeh.
\newblock The emergence of classical properties through interaction with the
  environment.
\newblock \emph{Z. Phys. B: Condensed Matter}, 59: 223--243, 1985.

\bibitem{zurek:03}
W.~H. Zurek.
\newblock Decoherence, einselection, and the quantum origins of the classical.
\newblock {\em Rev. Mod. Phys.}, 75:715--775, 2003.
\newblock arXiv:quant-ph/0105127v3.

\bibitem{schloss:07}
M.~Schlosshauer.
\newblock \emph{Decoherence and the Quantum to Classical Transition}.
\newblock Springer, Berlin, 2007.

\bibitem{bartlett:07}
S.~D. Bartlett, T.~Rudolph, and R.~W. Spekkens.
\newblock Reference frames, superselection rules, and quantum information.
\newblock {\em Rev. Mod. Phys.}, 79:555--609, 2007.
\newblock arXiv:quant-ph/0610030v3.

\bibitem{fuchs:10}
C.~Fuchs
\newblock QBism: The perimeter of quantum Bayesianism.
\newblock Preprint arXiv:1003.5209v1 [quant-ph], 2010.

\bibitem{zurek:04}
H.~Ollivier, D.~Poulin, and W.~H. Zurek.
\newblock Objective properties from subjective quantum states: Environment as a
  witness.
\newblock \emph{Phys. Rev. Lett.}, 93: 220401, 2004.
\newblock arXiv:quant-ph/0307229v2.

\bibitem{zurek:05}
H.~Ollivier, D.~Poulin, and W.~H. Zurek.
\newblock Environment as a witness: Selective proliferation of information and
  emergence of objectivity in a quantum universe.
\newblock \emph{Phys. Rev. A}, 72: 042113, 2005.
\newblock arXiv:quant-ph/0408125v3.

\bibitem{sandini:07}
G.~Sandini, G.~Metta, and D.~Vernon.
\newblock The iCub cognitive humanoid robot: An open-system research platform for enactive cognition.
\newblock In M. Lungarella, F. Iida, J. Bongard and R. Pfeifer, editors, {\em 50 Years of Artificial       Intelligence: Lecture Notes in Computer Science, Vol. 4850}, pages 358-369. Springer, Berlin, 2007.

\bibitem{zurek:09}
W.~H. Zurek.
\newblock Quantum Darwinism.
\newblock {\em Nat. Phys.}, 5:181--188, 2009.
\newblock (arXiv:0903.5082v1 [quant-ph]).

\bibitem{ashby:56}
W.~R. Ashby.
\newblock {\em An Introduction to Cybernetics}.
\newblock Chapman and Hall, London, 1956.

\bibitem{moore:56}
E.~F. Moore.
\newblock Gedankenexperiments on sequential machines.
\newblock In C.~W. Shannon and J.~McCarthy, editors, {\em Autonoma Studies},
  pages 129--155. Princeton University Press, Princeton, NJ, 1956.

\bibitem{fields:10}
C.~Fields.
\newblock Quantum Darwinism requires an extra-theoretical assumption of
  encoding redundancy.
\newblock {\em Int. J. Theor. Phys.}, 49:2523-2527, 2010.
\newblock arXiv:1003.5136v2 [quant-ph].

\bibitem{fields:11}
C.~Fields.
\newblock Classical system boundaries cannot be determined within quantum
  Darwinism.
\newblock {\em Phys. Essays}, 24:518-522, 2011.
\newblock arXiv:1008.0283v4 [quant-ph].

\bibitem{simons:08}
J.~S.~Simons, R.~N.~A.~Henson, S.~J.~Gilbert and P.~C.~Fletcher.
\newblock Separable forms of reality monitoring supported by anterior prefrontal cortex.
\newblock {\em J. Cognitive Neuroscience}, 20: 447-457, 2008.

\bibitem{fields:14}
C.~Fields
\newblock On the Ollivier-Poulin-Zurek definition of objectivity.
\newblock {\em Axiomathes}, 24: in press (DOI=10.1007/s10516-013-9218-3).
\newblock arxiv:1102.2826v5 [quant-ph].




\end{thebibliography}
\end{document}